# Interaction between Rydberg Excitons in Cuprous Oxide Revealed through Second Harmonic Generation

Dirk Semkat[1], Heinrich Stolz[2], Peter Grünwald[2], Andreas Farenbruch[1], Nikita V. Siverin[1], Dietmar Fröhlich[1], Dmitri R. Yakovlev[1,3], and Manfred Bayer[1,4]

July 23, 2025

[1]Experimentelle Physik 2, Technische Universität Dortmund, 44227 Dortmund, Germany,
[2]Institut für Physik, Universität Rostock, Albert-Einstein-Str. 23, 18059 Rostock, Germany,
[3]Ioffe Institute, Russian Academy of Sciences, 194021 St. Petersburg, Russia,
[4]Research Center FEMS, Technische Universität Dortmund, 44227 Dortmund, Germany.

**Abstract**

We report experimental and theoretical investigations of interacting excitons of the yellow series in cuprous oxide ($Cu_2O$) with principal quantum numbers up to $n = 7$ by means of second harmonic generation (SHG). Using picosecond pulsed laser excitation up to 10 GW/cm$^2$ peak intensity we observe a pronounced change of the spectra with increasing pump laser intensity: an energetic shift to lower absolute energies and a spectral broadening, while the absolute intensity for low powers scales with the square of the pump power, but saturates at higher powers. Concomitant with SHG we determined the density of the excitons excited by the ps pulse by measuring the two-photon absorption directly. This allows to derive quantitative values for the exciton-exciton interaction. Surprisingly, the results disagree both in magnitude and scaling with principal quantum number with those calculated by state-of-the art atomic-like van der Waals interaction theory. As a possible screening by an electron-hole plasma created by three-photon absorption into blue and violet band states could be ruled out, our results point toward fundamental differences between excitons and atoms.

## I.  Introduction

Since their first observation, Rydberg excitons in cuprous oxide ($Cu_2O$) [1] have turned out to be fascinating quantum objects with interesting properties ranging from quantum chaos [2] to strong Kerr nonlinearities [3] (for a review see [4]). Their similarity to Rydberg atoms [5] implies a strong dipole-dipole interaction leading to the phenomenon of Rydberg blockade [6]. Indeed, the strong changes in the one-photon absorption (OPA) spectrum with respect to the exciting laser power as observed in [1], was explained by this blockade effect. Recently, experiments demonstrated the *asymmetric blockade*, i.e., the change in the OPA spectrum of one Rydberg state due to the pumping of another different Rydberg state. The concomitant theoretical analysis showed that these results are consistent with an atomic like van der Waals interaction law for the excitons [7].

In all of these studies, the Rydberg excitons have been created by one-photon absorption in thin single crystalline platelets using continuous-wave (cw) excitation, preferentially by a single-frequency laser [1,8,9]. Due to the equal parity of the electronic bands from which the excitons originate, which are the $\Gamma_7^+$ valence and the $\Gamma_6^+$ conduction band [4], the strongest transitions belong to excitons with an $L=1$ envelope (P states) [10]. The interaction of the yellow 1S excitons with odd parity optical phonons results in a strong phonon sideband [11], which superimposes the P absorption lines and leads to their well-known asymmetry [12]. While this peculiarity does not affect experiments where by pump-probe schemes the interactions between Rydberg excitons have been investigated [1,7], another property of $Cu_2O$ has turned out to be quite problematic in these studies, the process of Auger-like scattering between two excitons [13,14]. In this process two yellow excitons, which can be both in different spin states, named conventionally ortho and para states [15], scatter at each other, whereby the energy of one of the excitons is transferred to the other resulting in an unbound electron-hole pair. On the one hand, this Auger process limits the maximum possible exciton density [15], on the other hand, it leads to the creation of an electron-hole plasma (EHP), even at very low excitation laser powers, whenever yellow 1S excitons are created via the phonon assisted absorption. This inevitably is the case in every scheme involving one-photon absorption [16]. In a typical experiment [1,5,13] application of a laser power of 100 µW would lead to EHP densities of the order of $10^{10}$ cm$^{-3}$ and plasma temperatures around 10 K [17,18]. The standard description for a low-density plasma, the Debye model [19], predicts that plasmas with such densities should lead to a negligible effect on Rydberg exciton states at principal quantum numbers below $n=20$ [17,20]. However, as shown recently, the Debye model is not applicable to such low temperature EHP [21,22] and even at very low concentrations an EHP leads to substantial effects on the absorption line shape of the P excitons and may mask completely the expected Rydberg blockade of Rydberg excitons [21,22,18]. So, in order to be able to unambiguously demonstrate the interaction between Rydberg excitons, one should look for scenarios in which the creation of an EHP made up from the yellow band states ("yellow" plasma) is suppressed. The rate models developed in [15,18] imply a significant delay for the creation of Auger-based EHP. Hence, one approach would be to switch from cw-excitation to subnanosecond pulses and to observe the changes in the absorption spectra of the P excitons with time. Another obvious way to avoid direct excitation of yellow 1S excitons is using the well-known two-photon absorption into S and D states with higher principal quantum number [23,24], as here any process involving odd parity optical phonons is symmetry forbidden. Utilizing as detection channel for the excitons the quadrupole emission one arrives at a process called "second harmonic generation" (SHG) [21]. Indeed, using femtosecond laser pulses it was shown recently that in this way exciton states with principal quantum numbers up to $n=9$ could be detected [25-27]. However, these studies have been focused on the physical mechanisms by which these processes become observable, i.e., the symmetry and polarization properties and on the influence of a magnetic field. Information on density dependent effects, i.e., on exciton-exciton interaction has not been obtained. This, however, would be highly desirable, as in such a scenario 1S excitons come up only via the decay of the Rydberg states, i.e., after times of the order of their lifetimes (see Fig. 1). As these are of the order of several picoseconds as has been established from their relaxation processes [28], and recently also by direct time-resolved spectroscopy [52], any influence of a yellow EHP that grows on much longer time scales can be avoided. These experiments also face the problem that due to the large spectral bandwidth of the femtosecond laser pulses many Rydberg states are excited simultaneously leading to interference effects [26,27]. Reducing the spectral bandwidth by employing nanosecond laser pulses allowed to observe well-resolved Rydberg states in SHG with principal quantum numbers up to $n=12$ [29]. However, no effects of an exciton-exciton interaction could be detected, because due to the rather long pulse duration the maximum laser power that could be applied before thermal heating of the sample sets in was far too low.

In this paper, we apply SHG using picosecond laser pulses with intensities up to the GW/cm$^2$ range tuned into resonance, i.e., twice the central pulse laser frequency, with the yellow exciton states expecting large enough exciton densities to study the interaction between excitons, which becomes visible in the quadrupolar emission (see Fig.1). Indeed, in our experiments we were able to observe SHG by exciton states with principal quantum numbers up to $n=9$ as in previous studies [26,27] but showing large non-linear effects like line shifts and saturation pointing toward very high densities. The reduced spectral width ($\approx 0.92$ meV compared to 10 meV for the femtosecond pulses) allowed us quantitative investigations of states up to $n=7$ with little disturbance by other exciton states. Note that the spectral width is still too large to avoid simultaneous excitation of Rydberg states with $n>7$ resulting in interference effects of the SHG signals and thus hindering a quantitative analysis. Furthermore, it turned out that there is the possibility of three-photon absorption (3PA) [30] into the blue and violet continuum producing on the time scale of the pump pulses free electrons and holes in these bands (called "blue-violet" EHP, see Fig. 1). This will directly interact with the yellow Rydberg excitons. As this process will increase with the third power of the pump intensity, we can separate the effect of such a blue-violet EHP on the Rydberg exciton states from the direct exciton-exciton interaction, which should depend on the exciton density, i.e., on the second power of the pump intensity.

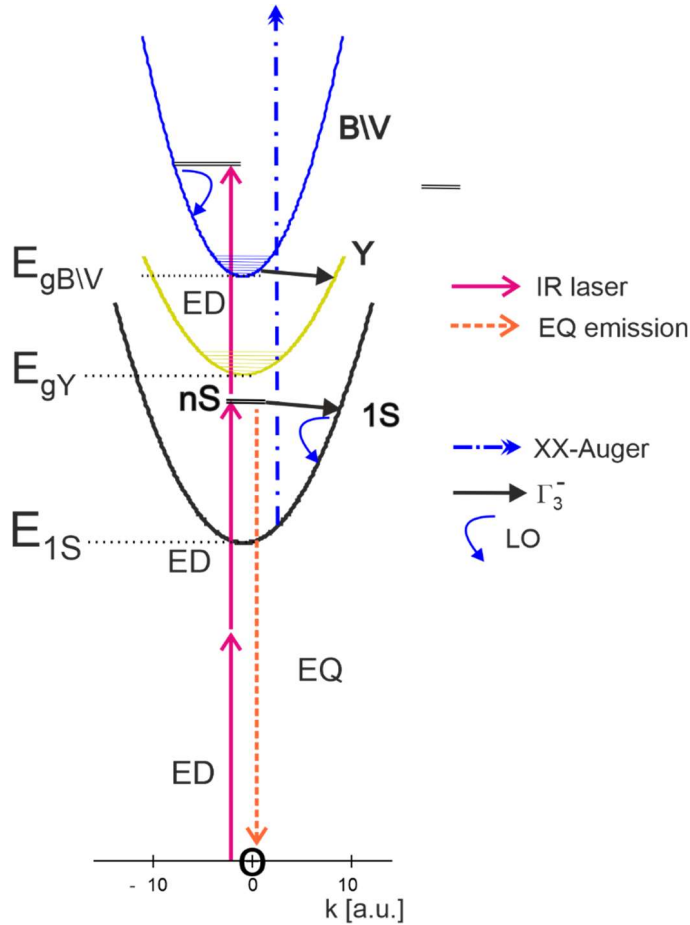

Fig. 1: Energy scheme of the important optical and relaxation process in Cu$_2$O in a two-particle picture.
O designates the crystal ground state, the parabolas the energy dispersion of the yellow 1S exciton (1S, black), the yellow band continuum (Y, yellow) and the blue/violet continuum (B\V, blue), the corresponding band gap energy, i.e. $E_{gY}, E_{gB\backslash V}$, is indicated by a dotted black line.
The black double lines denote the final states of (i) the two-photon (nS exciton) and of (ii) the three-photon absorption process. ED, EQ denote an electric dipole or quadrupole transition, respectively. The thick black arrows denote scattering by optical $\Gamma_3^-$ phonons, the light blue arrows by LO phonons. The blue dashed-dotted arrow indicates the bimolecular Auger-like process resulting in electron-hole pairs in high yellow and blue continuum states. The hatched areas near the bottom of the band continua denote electron-hole plasmas.

However, for a quantitative determination of the exciton-exciton interaction strength that would allow for a stringed test of any theory is the independent determination of the exciton density obtained in the experiment. Here, the method of SHG is also unique, as the number of absorbed photons of the IR laser, given by the loss of the laser energy in the sample, is exactly half of the number of excitons during the pulse, thus allowing to derive the two-photon absorption (TPA) coefficient.

Compared to previous studies addressing the range of principal quantum numbers above 10 [1,7], we extend here the experiments to principal quantum numbers below 10 and to much higher exciton densities, partly necessary to observe pronounced exciton-exciton interaction effects. The results obtained should allow a quantitative comparison with results of the atomic-like van der Waals calculations performed recently [5,7].

The paper is organized as follows. In Section II, we describe shortly the experimental setup and exemplarily present the results for both SHG and TPA measurements. In the next section we show the results of the analysis of the spectra by fitting Lorentzian-type line shapes and deriving in this way the dependence of total intensity, line shift and line width on pump power. In section IV we derive scaling laws and discuss the results within a Rydberg blockade model. Appendices discusses the effects of a blue-violet EHP, details of the TPA measusments. The paper closes with conclusions and an outlook.

## II. Experimental Methods and Results

Experimentally, SHG is measured by using a tunable pulse laser source with pulse width of about 3 ps and a spectral width of 0.92 meV (see Fig. 2). The repetition rate was set to 30 kHz, and the average laser power $P_L$ could be varied from 1 mW to 100 mW. The laser beam is focused to about 120 μm on the crystal sample immersed in superfluid helium at $T = 1.4\,\text{K}$. The thickness of the well-polished sample is $(29 \pm 1)$ μm as determined by interferometry. The excitation and detection polarization are chosen in such a way that the SHG is allowed for the even parity states (for details see Refs. 26,27). We then measure the spectra (spectral

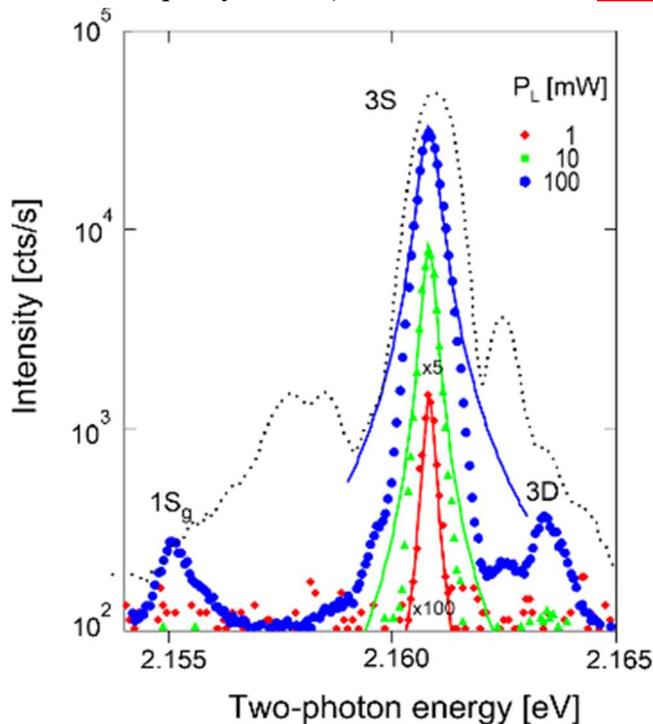

Fig. 2: SHG spectra (i.e., emitted light intensity vs. photon energy after subtraction of the dark noise of the camera) for excitation around the 3S yellow exciton for pump powers as indicated. The full colored lines are a line-shape fit with a Lorentzian (fit quality parameter $\sigma_{fit}$ between 0.03 and 0.05). Besides the dominant 3S line, we can identify weak transitions involving the 1S green exciton and a 3D yellow state. The dotted line gives the SHG of the laser itself (from a BBO crystal).

resolution $40\,\mu eV$) of the light emitted from the sample, spatially filtered so that only the light within the exciting laser beam is collected as function of laser photon energy and power. In all spectra a background due to the dark response of the photodetector is subtracted. To check whether 3PA processes lead to a creation of an EHP, we use an additional fs laser pulse, spectrally tuned to the absorption band of the Rydberg P excitons and transmitted through the sample, while creating simultaneously by a 3PA process with the ps laser pulses a blue-violet EHP

Typical results of the SHG measurements are shown in Fig. 2, where the laser pulse is centered at $2E_L = 2.161\,eV$, the 3S yellow exciton state. The SHG spectrum of the laser pulse itself was measured using a BBO crystal and is shown as the dotted line (note that the laser photon energy is always given as two times the infrared (IR) energy). The weak sidebands show that the laser pulse has a temporal phase structure, i.e., is not Fourier-limited [31]. Neglecting this, a simple analysis with a Gaussian line shape gives as half-width 0.92 meV. At low laser power (red points and line) we see a sharp emission spectrum with almost no background. With increasing laser power the intensity increases drastically (note the scaling factor of 20 between the red and green data). Concomitantly, the linewidth increases and sidebands show up, which can be assigned to other exciton states as the green 1S or a yellow 3D state.

By tuning the excitation laser pulse in resonance with the various exciton transitions we measured the SHG spectra for S lines from the $n=3$ to $n=7$ excitons as a function of laser power. The results are displayed in Fig. 3. We see that all resonances show a common behavior: at low excitation power ($P_L = 1\,mW$) the lines have a rather small linewidth (red curves). With increasing pump power the lines broaden and shift to lower energies. The magnitude of the shift strongly depends on the principal quantum number $n$, becoming quite large for the highest $n$. Also, the broadening of the lines at the highest powers is quite large (see the spectra at the highest power in Fig. 3c and d), which makes the analysis for the higher principal quantum numbers difficult. On the other hand, the richness of the observed features, i.e., excitation of S and D states with different quantum numbers, will open the door for interesting future experiments. These results are in complete contrast to those under cw excitation [1,7], where the effect of exciton- exciton interaction showed up only as more subtle changes in absorption such as bleaching of a line without considerable energy shift or broadening.

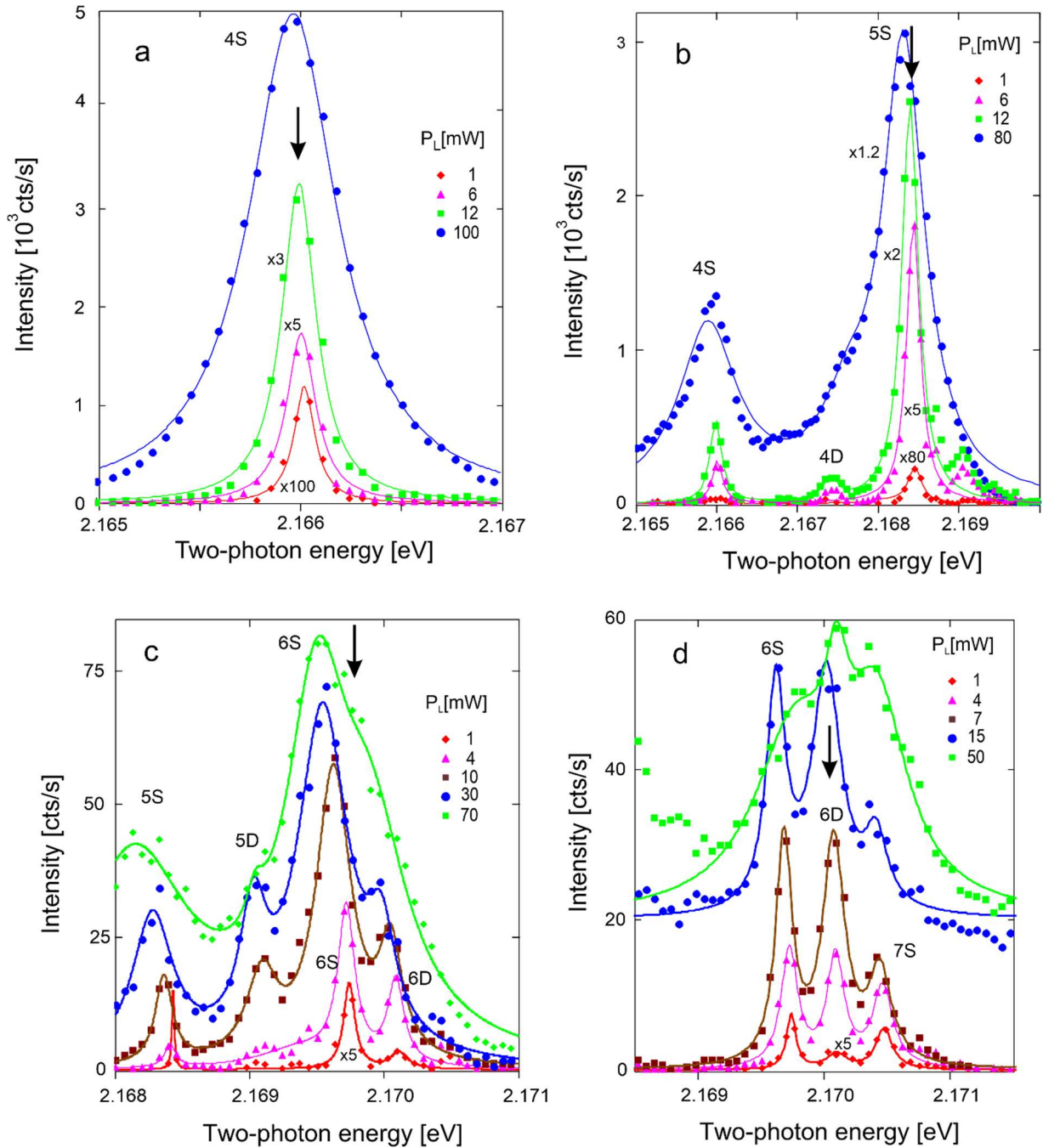

Fig. 3: SHG spectra for different laser central photon energies (indicated by the black arrow in each panel) and varying laser power as indicated. The exciton states involved are marked by *n*S and *n*D. The intensity scales are directly comparable for each series of measurements; scaling factors unequal to unity are indicated at each spectrum. In panel d, the spectra for 15 and 50 mW are shifted by an offset of 20 cts/s. Note that the fits for powers greater than 30 mW in panel c and d have a quality parameter $\sigma_{fit}$ (see below) larger than 0.1 and are not considered further.

## III. Data analysis

Quite generally, an emission line can be characterized by its first three moments, i.e., (i) the integrated intensity, (ii) the spectral position, and (iii) the line broadening. Assuming a Lorentzian lineshape for each of the SHG emission lines

$$L(E) = \frac{S_0}{\pi} \frac{\Gamma}{(E - E_X)^2 + \Gamma^2} \tag{1}$$

with $S_0$ denoting the peak area, $E_X$ the resonance energy and $\Gamma$ the phase relaxation rate, i.e. half of the full line width, allows one to fit the spectra almost quantitatively. In the following, we restrict the analysis to the S exciton lines. This allows a concise description of the effects of varying the pump power. Note that for the emission spectra there is no coupling to a phonon background as in case of the P excitons and the lines can be described by symmetric Lorentzians [32]. As seen in Figs. 2 and 3 this allows for a rather good fit, except at very high pump powers, where clear deviations are obvious (see Fig. 3a at $P_L = 100$ mW, where the wings of the 4S line underscores the Lorentzian shape).

To characterize the fit quality we use the following quality parameter

$$\sigma_{fit} = \sqrt{\frac{1}{N-1} \sum_{i=1}^{N} (Y_i - f_i)^2} \bigg/ \sum_{i=1}^{N} (Y_i) \tag{2}$$

with $Y$ denoting the background-free spectrum and $f$ the values of the fit function. We considered only fits with quality parameter < 0.1 as reliable.

The results for the peak area as function of pump power are shown graphically in Fig. 4, the results for the energy shifts and the line widths are shown in Fig. 5. From these results, one can deduce already some interesting phenomenological laws. The first one concerns the intensity of the SHG, which as a $\chi^{(2)}$ two-photon process should grow with the square of the pump power

$$\langle I_{SHG,n} \rangle = O_P(n) P_L^2 \:. \tag{3}$$

This is experimentally observed at low pump powers. The factor $O_P(n)$ should depend on the principal quantum number and is found to scale as $1/n^{3.35 \pm 0.1}$ for the S states (see Fig. 6a). However, at some critical power $P_c$ the increase is saturating and can be described (see Fig. 4, dashed lines) by a simple saturation function giving instead of Eq. (3)

$$\langle I_{SHG,n}\rangle = O_P(n)\frac{P_L^2}{1+P_L^2/P_c^2} \quad . \tag{4}$$

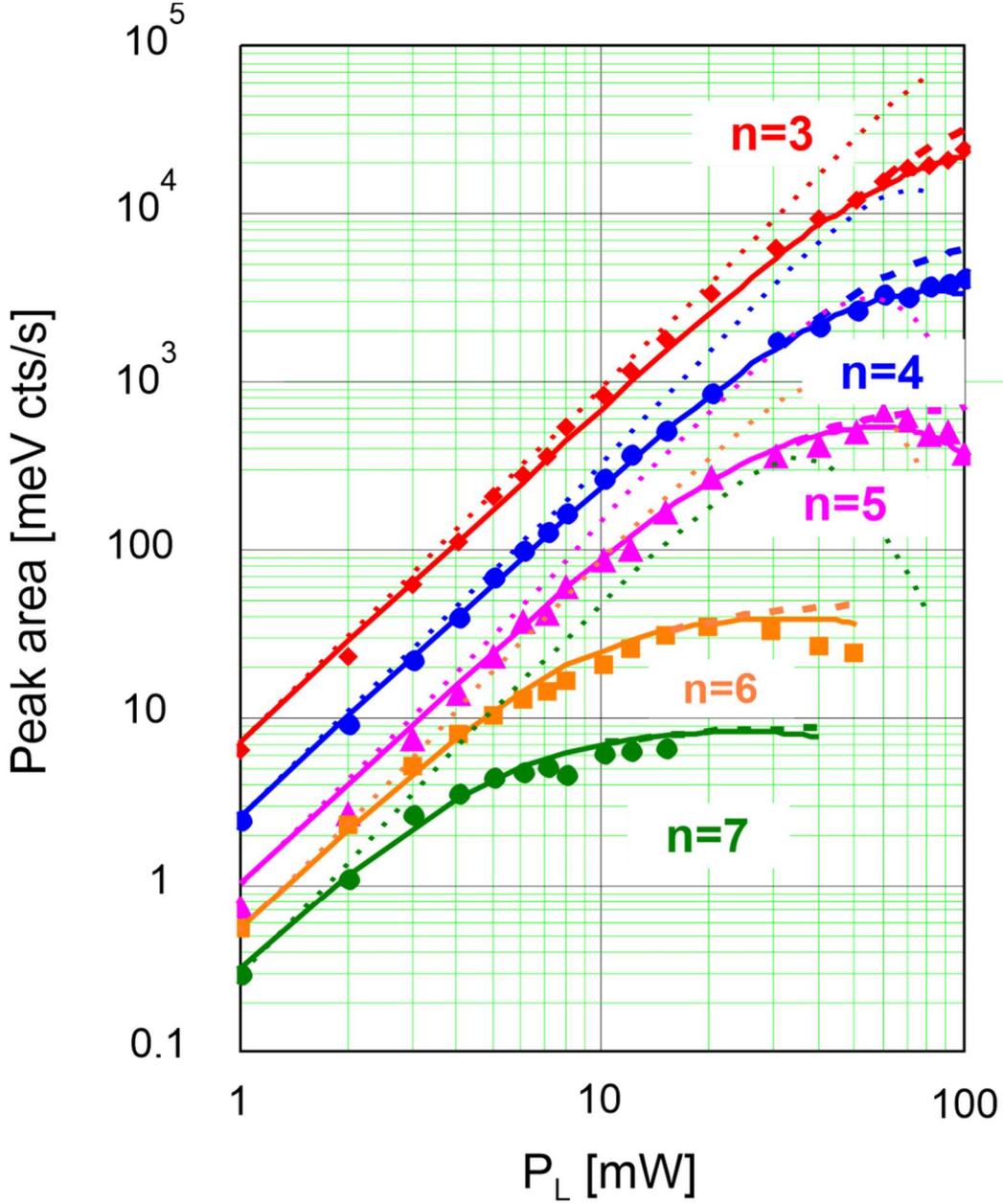

Fig. 4: Total SHG intensity of the S lines compared with the different theoretical models discussed in the text. The dotted lines give the results if only the effect of an EHP generated by the 3PA, the dashed lines those if only Rydberg blockade or equivalently a saturation law like Eq. (1.3) is considered. The full lines show the results if both mechanisms are acting. The linear increase with the square of the pump power at low powers is common to all models.

The critical power depends also on the principal quantum number of the states and is shown in Fig. 6a by the blue triangles. Interestingly, it shows no simple scaling behavior, which already points to a superposition of at least two saturation mechanisms. Also, the absolute reduction of intensity observed for $n = 5$ and $6$ at highest powers points to such a more complex mechanism.

Looking at the linewidths (Fig. 5b), we see a systematic increase with pump power. It can be fitted with the empirical law

$$\Gamma_2(n, P_L) = \Gamma_1(n)/2 + a_\Gamma(n) P_L^{2/3} . \tag{5}$$

The proportionality constant $a_\Gamma(n)$ is called the phase relaxation sensitivity, and the power dependence reflects that the effect should scale with the third root of the exciton density. The power independent constants $\Gamma_1(n)$ represent the energy relaxation rate or the inverse of the exciton relaxation time $T_1(n) = \hbar/\Gamma_1(n)$. It is plotted in Fig. 6c (red squares). As for P excitons, we expect it to be determined mainly by phonon relaxation into the 1S yellow exciton [28]. From the analysis of the linewidths determined from OPA spectra with applied electric field [32,33] a relation $\Gamma_1(n) = 7.0$ meV/$n^3$ (equivalent to $T_1(n) = 94\text{fs} \cdot n^3$) has been obtained [34]. This dependence is plotted in Fig. 6c (full red line). We see quite good agreement for low principal quantum numbers. The deviation at higher $n$ can be explained by the finite spectral resolution of the spectrometer of $40\,\mu\text{eV}$ [26] as seen by the magenta dashed line. The origin of the additional, power dependent contribution might be related to excitation induced dephasing [35]. Note that it has the effect that the total linewidth increases with principal quantum number even at low pump powers. The dependence $a_\Gamma(n)$ is shown in Fig. 6c (blue triangles) and shows a scaling $\propto n^{2.5}$ as seen by the straight line. The derivation of these relations from a many-body theory, however, lies outside the scope of this paper.

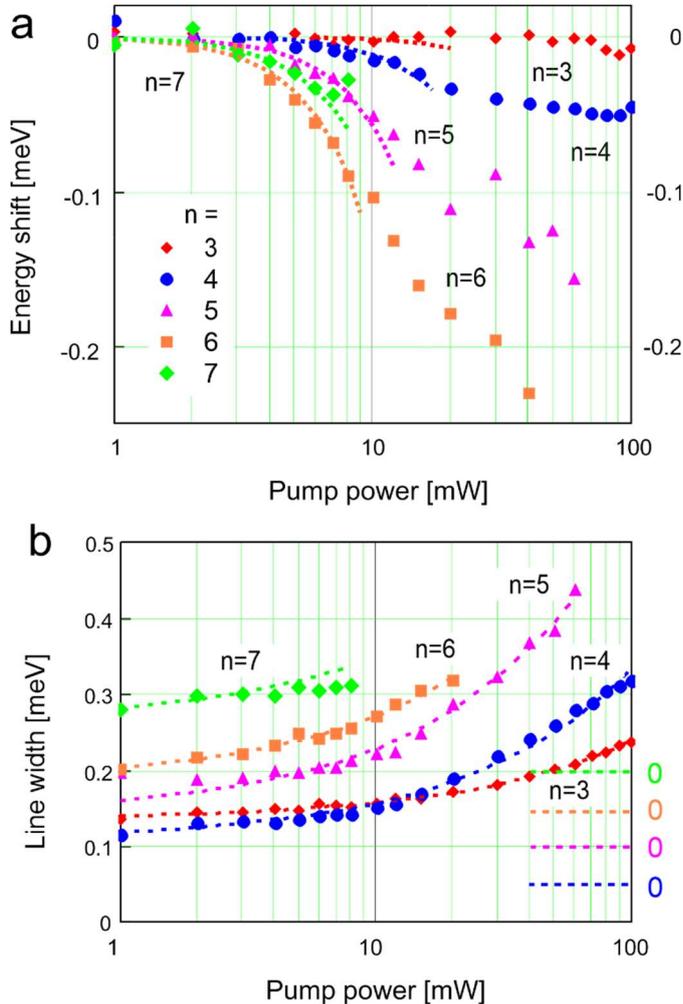

Fig. 5: Energy shift of the S resonances (panel a) and line width (panel b) as obtained from the fits of the SHG spectra. The experimental data (symbols) are plotted versus pump power (logarithmic scale) for states with $n = 3$ to $n = 7$. The dotted lines in panel a show the initial linear variation with the square of the pump power as given by Eq. (6). The slopes $S_{\Delta E}(n)$ are plotted in Fig. 6b as magenta squares. In panel b the results for different $n$ are shifted as indicated by the dashed lines marked on the right hand side with zeros. The dashed lines through the points are fits with Eq. (5), the phase-relaxation sensitivity $a_\Gamma(n)$ is shown also in Fig. 6b by the blue triangles. The quality of the fits is quite good (quality factor < 0.06).

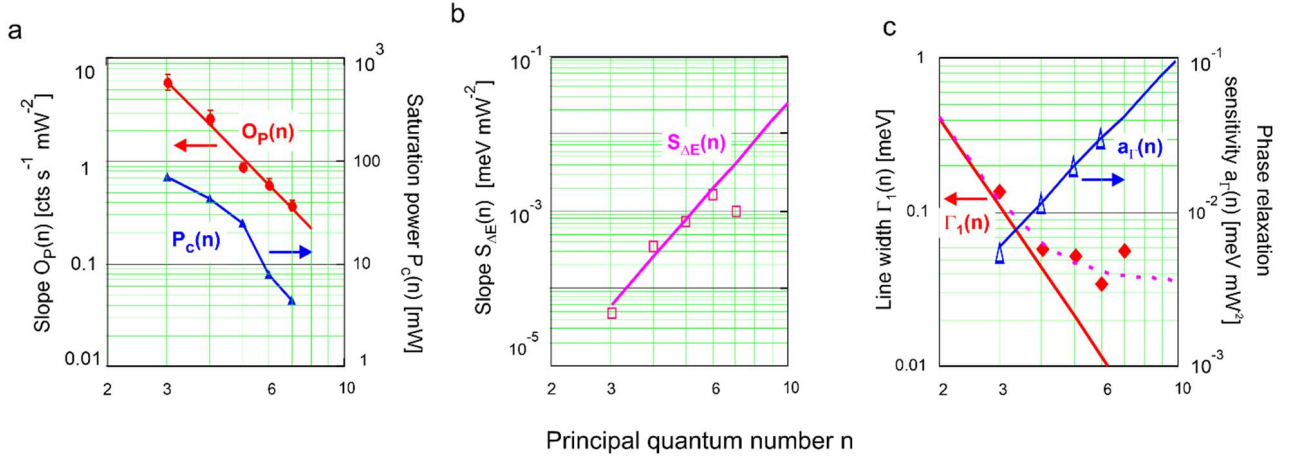

Fig. 6: Dependence of phenomenological fitting parameters $O_P$, $P_c$, $S_{\Delta E}$, $a_\Gamma$ and $\Gamma_1$ on principal quantum number $n$. Panel a shows the initial slope $O_P(n)$ of the total intensity at low pump power on principal quantum number $n$ for S states (red points). The full red line gives a dependence $\propto 1/n^{3.35\pm 0.1}$. The blue triangles give the dependence of the saturation power $P_c$ (Eq. (4)) on $n$ (right abscissa). The blue line is only a guide to the eye, the overall trend would correspond to a dependence $\propto 1/n^{3.5}$. Panel b shows the slope of the energy shift $S_{\Delta E}$ (magenta squares) on principal quantum number $n$, while panel c shows the phase relaxation sensitivity $a_\Gamma$ (blue triangles) and the lifetime $\Gamma_1$ (red diamonds). The full magenta line in panel b gives a dependence $S_{\Delta E} = 2.32 \cdot 10^{-7} n^{5.04\pm 0.1}$ meV mW$^{-2}$, while the blue line in panel c is the dependence $a_\Gamma(n) = 3.46 \cdot 10^{-4} n^{2.46\pm 0.1}$ meVmW$^{-2/3}$. The full red line in panel c gives the theoretical expected dependence $\Gamma_1(n) = 7.0\,\text{meV}/n^3$, while the magenta dashed line gives the result of a convolution of theoretical dependence with a constant additional broadening of $40\,\mu\text{eV}$.

Finally, let us look at the shift of the resonance energies for the S states. From Fig. 5a we already see that the shifts depend strongly on the principal quantum number and tend to saturate for high pump powers. Most interesting is that for low pump powers the shift grows with the square of the pump power, i.e., linear with the initial density of the excited excitons as given by Eq. (3). The dotted lines are representing the equation

$$\Delta E(n, P_L) = -S_{\Delta E}(n) P_L^2 \ . \tag{6}$$

The slopes $S_{\Delta E}(n)$ are plotted in Fig. 6b (open squares) as function of principal quantum number $n$ and show a clear dependence on the fifth power of $n$ except for $n = 7$, where the slope is reduced by a factor of 3.6 from the expected one. This might point to higher order effects in the interaction.

We can definitely exclude heating of the sample by the infrared pump laser, as the sample is immersed in superfluid helium with excellent cooling properties and any heating must show up in an energetic shift of all exciton lines by the same amount due to the shift of the band gap, which we do not observe (see Fig. 5a).

## IV. Discussion

### A. Intensity of the SHG process

Second harmonic generation has been studied previously in $Cu_2O$ [23-27,29]. Here, SHG has been considered as a process involving excitonic polarizations as light emitting dipoles, either in a semi-classical [27] or a simple quantum-optical [29] model, but neglecting completely the exciton-exciton interaction. Now in the concrete experiment, one excites with a very short laser pulse the excitons in a given state $n$ with subsequent much longer decay, but at such densities that we expect the exciton-exciton interaction to change the spectrum of the emitted light. As a fully time-dependent quantum-optical description of such a system is still an open problem and beyond the scope of this paper, we will apply the incoherent picture of *absorption followed by emission*, which is a good approximation for very short excitation pulses [36]. Here the emission process at a time $t$ is not influenced by the absorption and just given by the emission spectrum of an exciton system with the exciton density at time $t$. Since the two-photon absorption as the first step of SHG scales as the exciton wave function at $r = 0$ [37], in the hydrogen model this gives a proportionality $\propto 1/n^3$ [24]. However, according to recent calculations by Schweiner *et al.* [38], the oscillator strengths of the S states, and thus the wave functions at $r = 0$, scale as $\propto 1/n^{3.25}$. Therefore, we assume for the initial exciton density (at $t = 0$) the relation

$$\rho_0 = C_{PS} \frac{1}{n^{\beta_O}} (1 - f_{\text{loss}}) P_L^2 , \qquad (7)$$

with $\beta_O = 3.25$ and a constant $C_{PS}$, which depends only on the two-photon absorption constant and the excitation geometry. From the determination of the absolute exciton density for the 3S state (see Appendix B) we determined $C_{PS} = \left(135^{+30}_{-15}\right)$ mW$^{-2}$μm$^{-3}$, a value that gives for a laser power of 1 mW reasonable exciton densities between $5.8\,\mu m^{-3}$ and $0.57\,\mu m^{-3}$ for the 3S and the 6S exciton, respectively.

Anticipating the effects of an electron-hole plasma (see Appendix A) we included a factor $f_{loss}$ which gives the loss of oscillator strength due to a possible screening. However, for the higher $n$ states we should consider the effect of saturation, which reduces the initial density, as the critical power for, e.g., $n=6$ is only 8 mW. This can be done by transforming the saturation law Eq. (1.3) into a relation for the exciton density by introducing a saturation density in form of its inverse, which we identify with a blockade volume $V_{BL}$ resulting in the simple relation

$$\rho = \frac{\rho_0}{1+V_{BL}\rho_0} . \qquad (8)$$

The emission intensity of these excitons then is simply given by (with the probability of the radiative process $w_{rad}$)

$$S_X(P_L) \propto \rho \cdot w_{rad} = \frac{C_{PS}\frac{1}{n^{\beta_O}}(1-f_{loss})^2 P_L^2}{1+C_{PS}\frac{1}{n^{\beta_O}}(1-f_{loss})^2 P_L^2 V_{BL}} \frac{\Gamma_{rad}}{\Gamma_{tot}} \propto n^{3-2\beta_O}, \qquad (9)$$

We assume in the last step that the excitons radiate by the quadrupole process with radiative rate $\Gamma_{rad,n} \propto (1-f_{loss})/n^{\beta_O}$ as this is also proportional to the quadrupole oscillator strength and undergoes the same loss of oscillator strength by an EHP. We further assume that $\Gamma_{tot}$ is given by $\Gamma_1$. The resulting exponent of $-3.4$ agrees nicely with the experimental observations. Before going on in the analysis we discuss the implications of the exciton densities present in the experiments. As is well-known from atomic physics, there are two regions for the description of interatomic interactions, which are separated by the so-called Le Roy radius $R_{LR}(n,l) = 4(<r^2>)^{1/2}$ [39], where $<r^2>$ is the mean square of the electron-hole distance. For average distances larger than this radius, one can neglect higher order effects like multipole orders or exchange interactions and use a simple van der Waals description. This was recently confirmed also for excitons [5]. Using scaled hydrogen wave functions for the exciton states, we obtain $\sqrt{<r^2>} = 2n\sqrt{5n^2 - 1 - 3(L+1)}a_B$ with the exciton Bohr radius $a_B = 1.1$ nm [28]. In Fig. 7 we plot the equivalent Le Roy density, i.e., the density where the distance is equal to the Le Roy radius, as the dashed horizontal lines. Using the exciton densities from the analysis (Eq. (7), including the blockade as given by the blockade volumes from Fig. 9) as function of laser power we obtain the full lines. We see that at low enough laser power, these densities are smaller than the corresponding Le Roy densities, so that the usual van der Waals theory should be applicable in these low power ranges, e.g., for $n = 6$ up to 10 mW and up to 20 mW for $n = 3$. We therefore restricted the quantitative analysis with the following theory to this low power range. Details are given in Appendix C. While in this power range a linear dependence of the line shift with density has been found so that the restriction does not change the results, the saturation of the intensity is for low quantum numbers quite small, so that here we expect somewhat larger errors in the results as shown in Fig. 9. It should be noted that for the strong blockade effect observed at higher powers, the simple theory presented below is not valid.

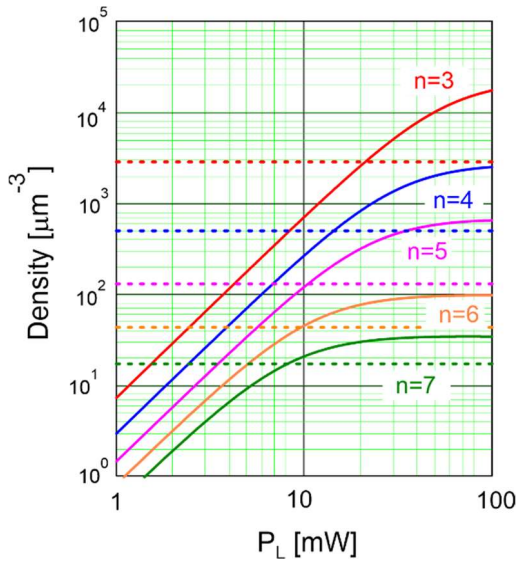

Fig. 7: Comparison of the mean exciton density (full colored lines) as function of pump power $P_L$ for $n = 3 - 7$ with the corresponding Le Roy density (i.e., the density where the distance is equal to the Le Roy radius) given by the dashed lines.

### B. Line shape and energy shift

The initially excited exciton density decays according to

$$\rho(t) = \rho_0 \exp(-t/T_1(n)) \tag{10}$$

with the energy relaxation time $T_1(n)$. We assume that this time does not depend on exciton density. We also neglect a possible influence of an EHP as it would scale in principle with the exciton oscillator strength [28] thus giving rise to a higher order effect. During this decay, the emission spectrum can be assumed to be a Lorentzian $L[E, E_X(n,\rho(t)), \Gamma_2(n,\rho(t))]$ with the density-dependent resonance energy $E_X(n,\rho) = E_X(n) + \Delta E(n,\rho)$. From quite general considerations of many-particle theory [19], one can expand the energy shift into a power series with respect to the density, which starts with the linear term

$$\Delta E(n,\rho) = a_1(n)\rho + \ldots, \tag{11}$$

with an interaction strength coefficient $a_1(n)$. From the experimental results (Fig. 5a) we can already infer that this assumption is valid only for low pump powers.

The width is determined by the total phase relaxation time $T_2(n,\rho) = \hbar/\Gamma_2(n,\rho)$ which comprises both pure dephasing processes and a possible inhomogeneous broadening, e.g., due to interaction induced energy shifts. For the total dephasing broadening we assume a density dependence of the form

$$\Gamma_2(n,\rho) = \Gamma_1(n)/2 + c_\Gamma(n)\rho^{\alpha_\Gamma} , \qquad (12)$$

where $c_\Gamma$ is obtained by combining Eq. (5) and Eq. (7). From this we expect the exponent to be $\alpha_\Gamma = 1/3$, which indeed is the case.

Then the measured SHG spectrum for excitons with principal quantum number $n$ is given by

$$S_n(E) \propto \Gamma_{rad,n} \int_0^\infty \rho(t) L[E, E_X(n,\rho(t)), \Gamma_2(n,\rho(t))] dt . \qquad (13)$$

Note that the line shape of the SHG emission will become asymmetric for large densities, but this is not relevant in the power range used for the analysis of the experiments, where we indeed did not observe this effect. The spectra obtained in this way are then fitted in the same way as the experimental spectra to a Lorentzian (Eq. (1)) giving averaged model quantities for the total intensity (which agree with those from section III), broadening and line shifts.

As an example of the theoretical results we show in Fig. 8 the calculated change of the average emission spectrum in case of the 5S exciton line, which obviously is able to describe the experimental findings.

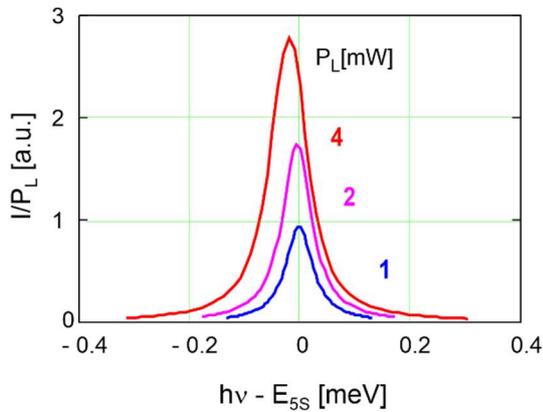

Fig. 8: Calculation of the power dependence of the SHG spectrum for the 5S state. Plotted is the spectrum for a given power (given by the numbers) normalized to the power.

Restricting again the analysis to the low power range and also neglecing any loss of oscillator strength, i.e., the effect of any blue-violet EHP and any saturation (see Appendix C for details), we obtain the dependence of the energy shift with density (Eq. (11)) for principal quantum numbers from 3 to 7. The slopes $a_1(n)$ obtained in this way are plotted in Fig. 9a as red diamonds and show a dependence

$$a_1(n) = -(3.47 \pm 0.5) \cdot 10^{-9} n^{(7.98 \pm 0.2)} \text{meV} \mu\text{m}^3 , \qquad (14)$$

except for $n = 7$, where the slope is reduced by a factor of 3.6. Note that due to the time averaging the behavior is somewhat different from the simple phenomenological fit (Eq. (6)).

The blockade volume obtained by the fit is shown in Fig. 9b by the blue triangles. While the states for $n = 3-5$ (full symbols) follow quite accurately a unique scaling law, for the states with $n = 6$ and 7 (open symbols) much larger blockade volumes are obtained. To explain this difference, we have to look at the actual spectral measurements (Fig. 3). Here we see that at larger $n$ in the experiments with higher applied pump powers other exciton species become dominating. Then one has to consider besides the self-blockade also the asymmetric blockade [5]. From the intensities one can estimate that for the 5S data the 5S exciton density is about 90% of the total concentration, while for the 6S measurements the 6S excitons comprise only about 36% of the total concentration. For the $n = 7$ experiments the fraction of 7S excitons is only 20%. Assuming a simple additivity of the blockade effects, this means that for $n = 6$ one overestimates the blockade by a factor of 3, while for $n = 7$ by a factor of five. Correcting the blockade volumes appropriately, we get the full symbols, which then follow the same scaling law

$$V_{BL}(n) = (5.88 \pm 0.5) \cdot 10^{-9} n^{(7.93 \pm 0.3)} \mu\text{m}^3 , \qquad (15)$$

Note that for the exciton-exciton interaction the situation is not so simple as it depends strongly on the quantum numbers [5] and is either repulsive or attractive. Unfortunately, there is neither for atomic systems nor excitons a theoretical estimation for the interaction between S and D states, so we can only speculate that the experimentally found reduction in $a_1(7)$ is due to such an effect.

### C. Possible influence of an electron-hole plasma

Due to the high intensities used for SHG, we inevitably create via 3PA free electrons in the $\Gamma_8^-$ conduction band and holes in the $\Gamma_{7,8}^+$ valence bands forming a blue-violet plasma, which interacts with the Rydberg states. As explained in detail in Appendix A we can deduce from the transmission experiments of the P excitons under pumping with 10 mW of IR laser pulses an initial density of the blue-violet plasma of about $1 \cdot 10^{11} \text{cm}^{-3}$ and a plasma temperature of about 25 K. As the density should scale with the third power of the pump power, we can calculate the expected plasma density in each case. Applying the theory of the plasma effect developed in the appendix this would lead to a reduction of oscillator strength for the 3S state at a power of 1 mW of about 0.01% and for the 6S state of 0.2%. Note that at 100 mW this would be much larger (12% and 80%, respectively). In the above analysis of the experiments, this effect has been included. As can be seen by the dotted lines in Fig. 4 it is of negligible importance in the low power range.

### D.  A simple Rydberg blockade mechanism

In this section we try to explain the observations of the saturation of exciton density by a blockade effect and the concomitant energy shift by the mechanism of a Rydberg blockade of excitons. Its idea can be formulated in a simple way [1,7,20]: If one creates an exciton at some point $\mathbf{r}$ then due to the exciton-exciton interaction potential $V(\mathbf{r}-\mathbf{r}')$ the energy of an exciton at point $\mathbf{r}'$ is shifted. If the resonance lies outside of half the laser bandwidth $\Delta E_L$ then the absorption process becomes suppressed. This happens below a critical distance $R_{BL}$. Assuming a simple power law for the interaction

$$V(\mathbf{r}-\mathbf{r}')=\frac{C(n)}{|\mathbf{r}-\mathbf{r}'|^p} \;, \tag{16}$$

with $C(n)$ being the interaction strength, the blockade distance is given by

$$R_{BL}(n)=\left(\frac{|C(n)|}{\Delta E_L/2+\Gamma_2(n)}\right)^{1/p} \;, \tag{17}$$

whereby we have to include the laser bandwidth as we have a spectrally broad pulse.

This blockade effect has two consequences:

(i)     Saturation of absorption

Due to the blockade a fraction of the crystal volume is not available for the exciton absorption process. The blockade volume per exciton is given by

$$V_{BL}(n)=\frac{4\pi}{3}R_{BL}(n)^3=\frac{4\pi}{3}\left(\frac{|C(n)|}{\Delta E_L/2+\Gamma_2(n)}\right)^{3/p} \;. \tag{18}$$

If the undisturbed exciton density as given by Eq. (7) would be $\rho_0$ then the density with blockade is given by Eq. (8). Since during their lifetime, e.g., an exciton with $n=6$ travels only a distance of about 50 nm, the decay of the excitons created initially does not change the spatial correlations.

(ii) Shift of the absorption spectrum

The average energy shift due to the interaction with an exciton ensemble with a constant density $\rho$ outside and zero inside the blockade volume is given by

$$\Delta E_X = 4\pi \rho \int_{R_c(n)}^{\infty} r^2 V(r) dr \quad . \tag{19}$$

Performing the integration gives (for $p > 3$)

$$\Delta E_X(n) = 4\pi \operatorname{sign}(C(n)) \frac{|C(n)|^{3/p} (\Delta E_L/2 + \Gamma_2(n))^{\frac{p-3}{p}}}{p-3} \rho \quad . \tag{20}$$

E. Comparison with the results of van der Waals theory

Inserting in Eqs. (18) and (20) the interaction strengths $C(n)$ calculated recently by applying the methods of atomic physics also to excitons [5], we can quantitatively compare the predictions of van der Waals theory shown as the magenta circles in Fig. 9 to our results. Firstly, and most important one can state that both theory and experiment agree surprisingly good, differing at most by a factor of three. However, one should note that both experiment and theory have their limitations. On the side of the theory, the calculation is based on approximate theoretical exciton energies and transition moments, which still differ in case of the energies considerably from experiment [54], whereby for the transition moments no experimental data exists at all. On the experimental side, the errors in the determination of the exciton densities are still quite large (see error bars in Fig. 9) and certainly need to be improved before final conclusions can be drawn.

Secondly, however, experiment and theory obviously differ in an important aspect, namely the scaling with the principal quantum number, which is not influenced by uncertainties in the exciton density. A closer look to the relevant equation reveal that both blockade volume and interaction strength coefficient $a_1(n)$ should scale with the same power of the principal quantum number, which indeed is the case for the experimental data (see Eqn. (14) and (15)). Using the value for $p = 6$ derived below, the scaling of the interaction coefficient derived from the experiment should be $15.9 \pm 0.3$, which is considerably larger than the value of 11 given by the van der Waals theory [5]. However, one should note that this is only valid in the limit of large principal quantum numbers $n$, while for small $n$ this depends on the detailed behavior of energy levels and transition moments. Of course, more clarity on this point would be possible if one could extend the experiments to higher principal quantum numbers, where the van der Waals theory is valid and the discrepancy would be more pronounced.

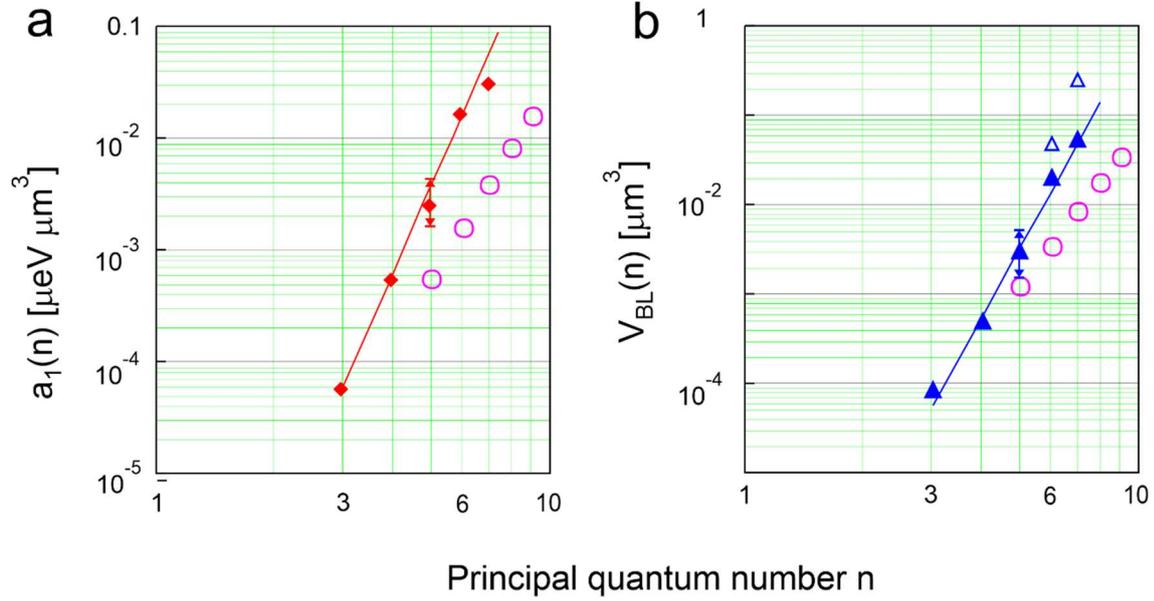

Fig. 9: Interaction strengths $a_1(n)$ (panel a) and blockade volume $V_{BL}(n)$ (panel b) vs. principal quantum number $n$.
Panel a: The filled diamonds show the results obtained by fitting the spectra for different pump powers as described in Appendix C, the open circles are calculated using the $C_6$ interaction constants (for $n \geq 5$) taken from Ref. 5. The full line is a power law $\propto n^{8.0}$ (Eq. (14)). Panel b: For $n \leq 5$ the full triangles show the results obtained directly by fitting the intensity as described in Appendix C, for $n=6$ and $n=7$ these results are given by the open symbols. Here the full symbols are corrected by taking the other excitons besides the relevant one into account as discussed in the text. The full line gives the scaling $\propto n^{7.9}$ (Eq. (15)). The predictions for $V_{BL}(n)$ by Eq. (18) of the van der Waals theory of exciton-exciton interaction are plotted as the magenta circles. The errors in the experimental data are indicated for $n=5$ as typical example and are mainly due to the uncertainty in the determination of the exciton density (see Appendix B).

Now taking the ratio of both quantities we obtain

$$V_{BL}(n)/a_1(n) = \frac{1}{3}\left(\frac{1}{\Delta E_L/2 + \Gamma_2(n)}\right)^{3/p+(p-3)/p}(p-3) = \frac{1}{3}\frac{p-3}{\Delta E_L/2 + \Gamma_2(n)}. \quad (21)$$

This allows to obtain the power $p$ of the interaction law independent of errors in the determination of the exciton density. Experimentally, the ratio is found to be $(1.59 \pm 0.2)\,\text{meV}^{-1}$. Using for the half-width of the laser 0.92 meV (see Fig. 2) and assuming an average width of the exciton states of about 0.15 meV one obtains $p = 6.1 \pm 0.3$, which agrees surprisingly well with $p = 6$ expected for an atomic-like exciton-exciton interaction based on an expansion of the full Coulomb interaction between the electron and hole [5]. However, to describe the observation at higher powers an interaction theory for distances smaller than the Le Roy radius is required. Whether this is still possible with an atomic-like description of the exciton-exciton interaction or requires a many-body theory as proposed in [37,38] is at present unclear and needs further investigations.

### V.   Conclusions

In this paper we have shown that by using SHG with ps laser pulses it is possible to study yellow excitons in $Cu_2O$ with principal quantum numbers up to $n = 7$ at such densities that clear non-linear effects become observable. These consist primarily in (i) an energy shift of the exciton resonance lines and (ii) a saturation of the intensity, but the experimental results reveal already a very rich scenario of non-linear effects like, e.g., multi-exciton interactions. The detailed analysis of these effects, however, would go beyond the aims of this paper. In the same vane we stress that the exciton densities achieved in the experiments at high pump powers imply more than one exciton within a volume of the size of an optical wavelength. Thus, we highly suspect cooperative radiative processes to occur in these power regions.

In the experimental scenario, the presence of a yellow EHP can be excluded, but the occurrence of a blue-violet plasma excited by three-photon absorption at high pump power has to be considered, but was shown to be ineffective in the power range of the experiments and therefore could not explain the non-linear effects. Hence, these can only stem from exciton-exciton interactions. The observed energy shift of the lines scales at low pump powers linearly with the exciton density, which could be determined from measuring the two-photon absorption directly. The analysis allows us to deduce the strength of the exciton-exciton interaction in the S-S scattering channel for principal quantum numbers from 3 to 7 quantitatively. As in this power range the mean exciton distance is larger than the Le Roy radius, the exciton-exciton interaction should be determined simply by an interaction law. This allows to compare our results with the standard van der Waals theory [5]. To this end, we put forward a Rydberg blockade model, from which we can deduce the line shifts and the saturation behavior given the interaction law of the excitons. While we find the interaction strength to be of the same order as predicted by theory the scaling with the principal quantum number is different from atomic-like van der Waals calculations [5]. We can rule out that this is due to too high exciton densities, as we restricted the analysis to low pump powers, where the van der Waals approximation should be valid. One possible reason may lie in deficiencies of the theory itself, as the hitherto used energy levels and transition dipole moments are derived from theoretical exciton models and not experiment. At higher exciton densities the inadequacy of a simple van der Waals description is already visible in our experiments in form of the observed nonlinearity of the energy shifts with power. This may point towards the claims of theoretical papers (see e.g. Refs. 40 and 41) that due to the composite fermion nature the interaction between excitons cannot be described by a potential function depending only on the distance of the center of masses of the excitons, but must be non-local. In either case, our results will provide a solid test for any theory of interacting excitons to be developed, e.g., along the lines of Ref. 42, including the origin of the observed excitation induced dephasing. On the experimental side in the future, it is urgently needed to extend the investigated exciton states to higher principal quantum numbers to overlap with other types of measurements. Progress here will require the use of laser pulses with much smaller line width, preferentially optimized to the linewidth of the Rydberg states to allow selective excitation of single states. And last but not least, the quantitative determination of exciton densities from two-photon absorption must be improved, which will be extremely challenging as very small effects of less than 0.1% absorbance are expected.

**Acknowledgements**

We thank Wolf-Dietrich Kraeft, University of Rostock, Germany, for helpful discussions. D.S. thanks the Deutsche Forschungsgemeinschaft for financial support (project number SE 2885/1-1), the Dortmund side acknowledges the support by the Deutsche Forschungsgemeinschaft through the International Collaborative Research Centre TRR 142 (Project A11). Finally, we would like to give special thanks to the referees. Their in-depth review and insightful commentaries led to a substantial improvement of the paper.

**Appendix A: Influence of a blue and violet electron-hole plasma on yellow exciton states**

In this appendix we will investigate the influence of an electron-hole plasma (EHP) possibly created by three-photon absorption (3PA) of the infrared pump laser. It is well known that in any semiconductor with a strong dipole-allowed band gap the optical absorption by a 3PA process is quite strong. For a simple two-band model, in Refs. 43 and 44 an expression is derived for the three-photon absorption coefficient $\alpha_3$ that via

$$\frac{d\rho_{eh}}{dt} = \alpha_3 I_{IR}^3 / 3\hbar\omega_{IR} , \tag{22}$$

determines the density of charge carriers $\rho_{eh}$ of the EHP, $\hbar\omega_{IR}$ is the photon energy of the infrared pump laser and $I_{IR}$ its intensity. It is shown that it depends only on the band-edge energy $E_g$, the momentum matrix element between the band states is given by the Kane energy $E_P = |P_{cv}|^2 / 2m_0$ (with $m_0$ the electron mass and $P_{cv}$ the momentum matrix element between valence and conduction band) and the index of refraction $n_b$ at the IR photon energy. The result is

$$\alpha_3(\hbar\omega_{IR}) = Q_3 \frac{E_P^{3/2}}{n_b^3 E_g^7} F_3\left(3\frac{\hbar\omega_{IR}}{E_g}\right), \text{ with } F_3(x) = \frac{(x-1)^{1/2}}{x^9} , \tag{23}$$

and

$$Q_3 = \frac{3^{10}\sqrt{2}}{8}\pi^2 \alpha_{SF}^{-3} \frac{\hbar^5}{\sqrt{2m_0}} = 5.0 \text{ eV}^{11/2}\text{cm}^3\text{GW}^{-2} , \tag{24}$$

where $\alpha_{SF}$ is the Sommerfeld fine-structure constant.

In Cu$_2$O the lowest dipole-active transition is that from the $\Gamma_7^+$ and $\Gamma_8^+$ valence bands to the $\Gamma_8^-$ conduction band (so called "blue" and "violet" transitions), whereby the violet transition is the strongest (ratio of oscillator strengths 5/3 [45]). It turns out, however, that we have to take both components into account, as they act differently on the yellow exciton states. Taking for the Kane energy of the violet transition $E_P \approx 25$ eV as in II-VI semiconductors, $n_b = 2.6$ and $E_{gV} = 2.765$ eV (violet band edge [45]), we obtain at $\hbar\omega_{IR} = 1.04$ eV a value $\alpha_{3V} = 3.96 \cdot 10^{-3}$ cm$^3$GW$^{-2}$. For 3PA in the blue states we would obtain, using as Kane energy $E_P \approx 15$ eV and the blue band gap ($E_{gB} = 2.635$ eV), an absorption coefficient of $\alpha_{3B} = 1.84 \cdot 10^{-3}$ cm$^3$GW$^{-2}$. However, one should be aware that due to the strong non-parabolicity of both the valence bands and the conduction band, the calculation can give only an order of magnitude for $\alpha_3$.

From the given experimental data (3 ps pulse width, repetition rate 30 kHz, average power 1 mW to 100 mW, focal spot size 120 μm), one obtains for 40 mW average power a peak intensity of 5.9 GW/cm$^2$. Neglecting decay this would give a violet electron-hole density of about $\rho_{ehV} = 4.9 \cdot 10^{15}$ cm$^{-3}$, whereas the blue plasma would have a concentration half of this. As the excess energy is around 450 meV, i.e., $3\hbar\omega_{IR} - E_{gB\backslash V}$ this is a rather hot plasma with a high density. However, by LO phonon scattering via Fröhlich interaction the electrons and holes relax very fast to the bottom of the band. For the LO$_2$ phonon (82.1 meV) this requires about 50 fs, while for the low energy LO$_1$ phonon the scattering time is about 300 fs [18]. So within the pulse duration of 3 ps the carriers relax down to the point, where the kinetic energy is not enough for LO scattering. The following acoustic scattering is much slower. Assuming equipartition of the excess energy there remains an energy between 0 and 19 meV (LO$_1$ phonon energy). So the plasma gets much cooler with a maximum temperature of about $T_{ehvio} = 200$ K (due to the non-parabolicity of the bands it is impossible to obtain more accurate values without enormous efforts which is beyond the scope of this paper). During relaxation the carriers in principle might scatter by phonon relaxation into the lower-lying $\Gamma_{6c}^+$ conduction band or the $\Gamma_7^+$ valence band forming a yellow plasma. The fact that the photoluminescence excitation spectroscopy of the yellow 1S excitons [9] shows at the spectral position of the blue and violet exciton states a minimum in the response reveals that the probability of such a scattering process must be quite small. Therefore, a yellow plasma does form on a much longer time-scale than the lifetime of the Rydberg excitons of several tens of picoseconds.

From previous studies of multi-band EHP [46-48] it is known that the direct and exchange Coulomb interactions between different band states are reduced due to the orthogonality of band states at the same wave vector. This has the consequence that the band-edge shift of an unoccupied band, like the $\Gamma_{7v}^+$ valence and the $\Gamma_{6c}^+$ conduction bands (yellow bands) is not influenced by an occupation of the violet band states. Only the blue states might have some influence due to the common $\Gamma_{7v}^+$ hole states. We expect that for this effect the existing many-particle theory [21] can be modified. Indeed, taking only a $\Gamma_{7v}^+$ hole plasma into account, the theory gives the same expression for the band-gap shift, but reduced by a factor of $\sqrt{2}$

$$\Delta E_{ybg}(\rho_{\Gamma_7^+}, T_{\Gamma_7^+}) = -0.49 \text{ meV} \left(\rho_{\Gamma_7^+} \cdot T_{\Gamma_7^+}\right)^{1/4}. \tag{25}$$

For the violet plasma, on the other hand, the only effect will be the screening of the Coulomb interaction between electron and hole of the yellow exciton. In a first approximation, we can assume that the Coulomb potential is replaced by a screened potential

$$V(r) = \frac{e_0^2}{4\pi\varepsilon_0\varepsilon_r} \exp(-\kappa_D r), \tag{26}$$

whereby we use the Debye screening wave vector as the plasma is nondegenerate

$$\kappa_D = \left(\frac{2\rho_{ehvio} e_0^2}{\varepsilon_0 \varepsilon_r k_B T_{ehvio}}\right)^{1/2}. \tag{27}$$

Here $\varepsilon_0, \varepsilon_r$ denote the dielectric permittivity of the vacuum and of the medium, $e_0$ the elementary charge, $k_B$ the Boltzmann constant and $\rho_{ehvio}, T_{ehvio}$ are the plasma density and temperature.

In the following, we use the parameter

$$\xi = \frac{a_B}{2\lambda_D} = \frac{1}{2} a_B \kappa_D. \tag{28}$$

with $a_B = 1.106$ nm being the yellow exciton Bohr radius [11]. For the blue-violet plasma one expects from the above calculation a plasma parameter $\xi_v \approx 0.0178$.

The solutions of the exciton Schrödinger equation for a screened Coulomb interaction are well known (see, e.g., Ref. 47). They can be most easily obtained by a variational procedure assuming a variable Bohr radius $a_B$ but keeping the principal quantum number in the hydrogen wave functions [50]. In the variational calculation one has to minimize the expression (with $x = a_B / a_{B0}$, $a_{B0}$ the undisturbed Bohr radius)

$$e(x,n,l,\xi) = \left(\frac{1}{x^2} - \frac{2}{x} \cdot \frac{1}{(1+n\xi x)^{2n}} \cdot {}_2F_1(-l-n, 1+l-n, 1, (n\xi x)^2)\right), \tag{29}$$

where ${}_2F_1(a,b,c,z)$ denotes the Gaussian hypergeometric function [53]. Denoting by $x_0(n,l,\xi)$ the extremum value of $x$, the oscillator strength according to Elliott [10] is then given as

$$f_{osc,P} = f_{osc,P0} \cdot x_0^{-5} \tag{30}$$

for P and

$$f_{osc,S} = f_{osc,0} \cdot x_0^{-3} \tag{31}$$

for S states.

We have found that the results of the variational procedure for the oscillator strength can be expressed very accurately by the formula

$$f_{osc}(n,\xi) / f_{osc}(n,0) = 1 - \frac{2}{\pi}\arctan\left(S_{osc}n^4\xi^2\right), \tag{32}$$

with $S_{osc}=13$ for the S and $S_{osc}=20$ for the P states (see Fig. A1).

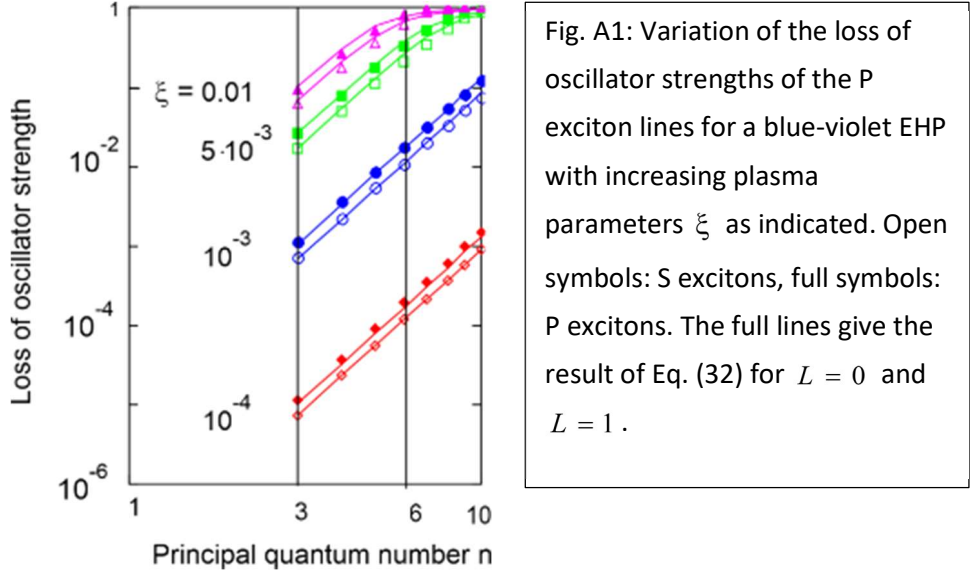

Fig. A1: Variation of the loss of oscillator strengths of the P exciton lines for a blue-violet EHP with increasing plasma parameters $\xi$ as indicated. Open symbols: S excitons, full symbols: P excitons. The full lines give the result of Eq. (32) for $L = 0$ and $L = 1$.

The problem with this model is that it predicts energy shifts of the lines which are quite large, all lines shifting at low plasma densities in the same way, which is well-known for the Debye plasma model [49,50]. Actually, in the experiments we did observe a trend for saturation of the line shifts at high pump powers, i.e., where the effect of an EHP would be strongest. However, a quantitative analysis would require in first place an accurate description of the higher order effects of the exciton-exciton interaction itself, which is not available. Secondly, we would need a true many-body description of the effect of a plasma of different band states on excitons. In lack of all these, we will use Eq. (32) to analyze the following pump-probe measurements.

In conclusion, we expect two different effects of a blue-violet plasma: (i) a band-edge shift due to the blue component given by Eq. (25), and (ii) a reduction of oscillator strength from the violet plasma given by Eq. (1.30). In Fig. A2 we show a typical result of the change of the one-photon transmission spectrum (panel a) when applying ps laser pulses (average power 40 mW) to a $Cu_2O$ sample with a thickness of 30 μm. Note that in case of a fs probe (pulse duration about 100 fs) we do not expect strong geometry-induced coherence effects, as the round trip time is more than 600 fs. So we can calculate from the transmission $T$ directly the absorption constant α using the Bouguert-Lambert law (see e.g. Ref. 16 for a discussion) and a reference spectrum of the laser pulse (see inset in Fig. A2a). The results are shown in Fig. A2b. Here the upper trace (red line) shows the absorption spectrum at zero pump power while the blue line shows the difference between the absorption spectrum at $P = 40$ mW and zero pump power. We see clearly the two expected effects: (i) a shift of the band edge of about 2.5 meV and (ii) a reduction of oscillator strength seen by the diminishing of the absorption lines. By fitting asymmetric Lorentzians to the P absorption lines together with the background (for a detailed description see Ref. 16) we obtained the peak areas $O(n)$ of the P lines, which are proportional to the oscillator strengths agreeing with those reported earlier [16].

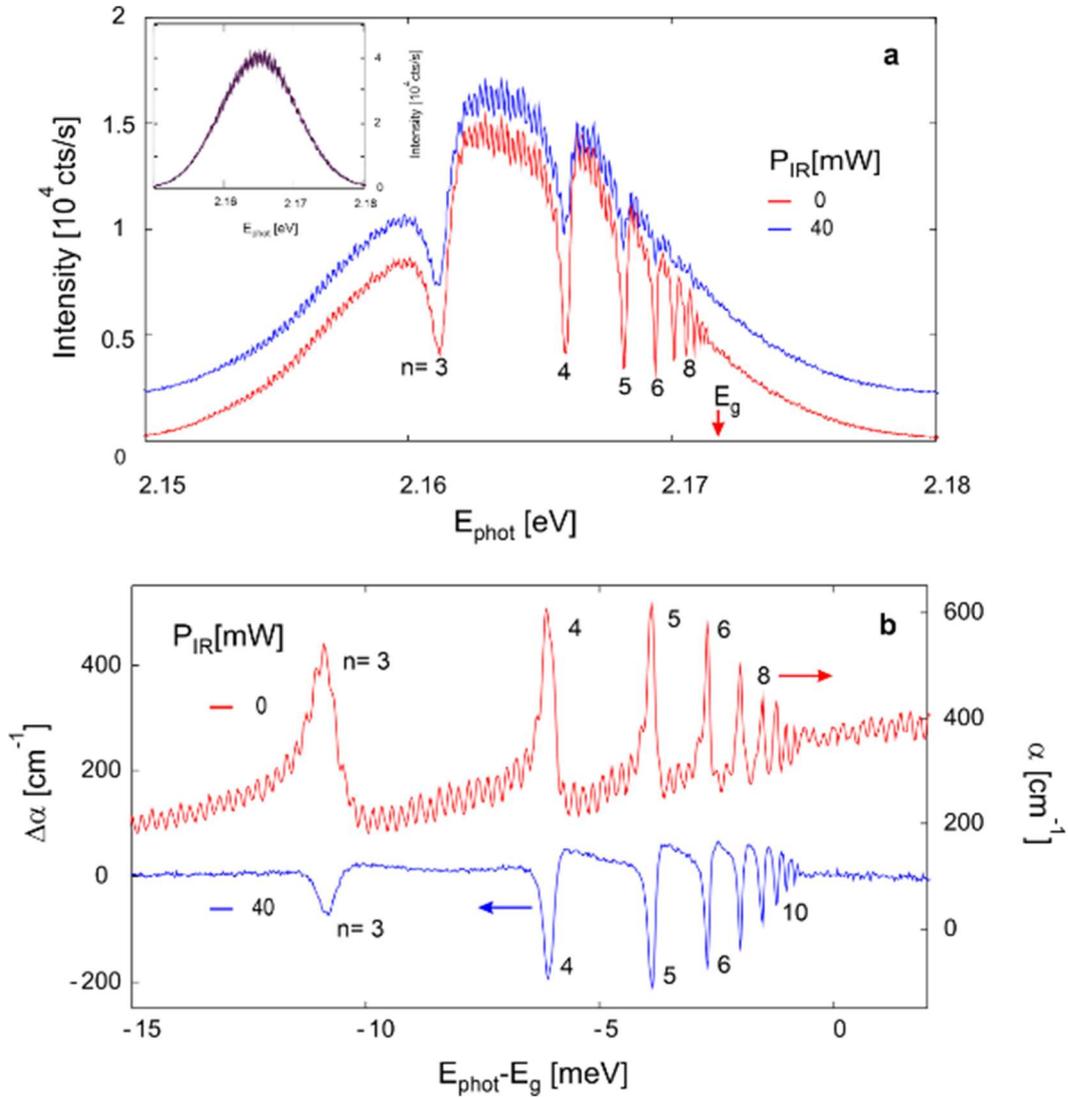

Fig. A2: In panel a the transmission spectra of a thin Cu$_2$O crystal are shown in the region of the yellow band edge (red arrow) without (red line) and with an infrared pump laser at 1.04 eV and an average power of 40 mW (blue line). In the spectrum at zero pump power, we observe P lines up to $n = 11$, whereas at a pump power of 40 mW the maximal observable principal quantum number decreases to $n = 6$. The inset shows the reference spectrum of the laser pulse. In panel b the upper curve (red line) shows the absorption spectrum calculated from the data in panel a as described in the text. The lower spectrum shows the difference of the absorption between zero and 40 mW pump power $\Delta\alpha = \alpha(P = 40 \text{ mW}) - \alpha(P = 0 \text{ mW})$. Note that the interference pattern that is visible in the spectra completely cancels in the difference spectrum making the analysis much cleaner.

To obtain the reduction of oscillator strength of the P lines due to the presence of a blue-violet EHP one then simply has to integrate the difference spectra over the spectral region around each $n$P line, which is chosen wide enough to include possible line shifts, resulting in

$$f_{osc}(n,P_{IR})/f_{osc}(n,0) = 1 - \frac{\int_n \Delta\alpha(E,P_{IR})dE}{O(n)} \quad . \tag{33}$$

The data obtained in this way are plotted in Fig. A3 and compared to the predictions of the simple Debye model. The agreement is quite good for a plasma parameter of $\xi = 0.0115$, which is in reasonable agreement with the predictions from the 3PA absorption above. If we use Eq. (28) this gives a screening wave vector $\kappa_D = 0.023/a_B$. Using Eq. (25), we have together with Eq. (27) two equations to determine both plasma density and temperature as $\rho_{ehvio} = (145\pm15)\cdot 10^{12}\,\text{cm}^{-3}$ and $T_{ehvio} = (10\pm2)\,\text{K}$. The electron and hole density thus determined are obviously much lower (factor $\approx 30$) than that estimated from the above calculation of the 3PA coefficient showing that the two-band model used is much too simple.

Now the 3PA should vary as the third power of the laser intensity. This means that the plasma parameter $\xi$, which scales with the square root of density, scales with laser power as $\xi \propto P_L^{3/2}$. Thus, at 10 mW it should be $\xi(10\,\text{mW}) = 1.4\cdot 10^{-3}$ and at 1 mW $\xi(1\,\text{mW}) = 4.5\cdot 10^{-5}$. Assuming the same temperature, this corresponds to plasma densities of about $1.0\cdot 10^{12}\,\text{cm}^{-3}$ and $1.0\cdot 10^{9}\,\text{cm}^{-3}$.

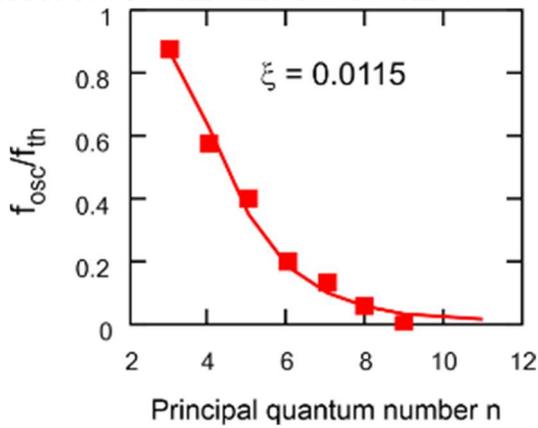

Fig. A3: Loss of P exciton oscillator strength due to a blue-violet plasma. The filled squares give the loss obtained from the experiments shown in Fig. A2, the full line depicts Eq. (32) for $\xi = 0.0115$.

If we want to apply the plasma model to the experimental situation of the SHG measurements we have to consider that (i) . Considering the S excitons, the loss of oscillator strength would be given by Eq. (32) with $S_{osc} = 13$. The reduction of oscillator strength for the above given plasma densities is shown in Fig. A4.

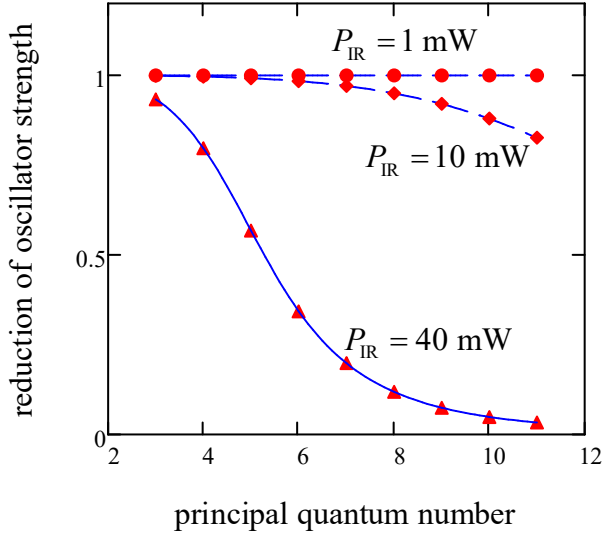

Fig. A4: Loss of S exciton oscillator strength due to a blue-violet plasma for various plasma densities expressed by the plasma parameter $\xi$. In the calculations we included the effect of a reduction of the absorption coefficient due to a change of pump laser photon energy to 1.08 eV, to $\alpha_3 = 3.0 \cdot 10^{-3}$ cm$^3$GW$^{-2}$, a 25% reduction. The triangles correspond to an IR laser power of 40 mW, the diamonds to 10 mW and the circles to 1 mW. The full lines are a guide to the eye.

Then the intensity of the SHG signal is determined by Eq. (13) whereby we include the loss effect both in absorption and emission, but neglect completely any blockade effect. We see from Fig. A4, that for powers below 10 mW, any effect of a 3PA-generated plasma can be safely neglected, while it must be considered for the higher pump powers. We therefore have to restrict the quantitative analysis of the data to the low power range.

## Appendix B: Determination of the exciton density from two-photon absorption

The exciton density can be determined by measuring the transmission loss of the IR pumping laser when its wavelength is in resonance with an exciton state. The setup is straightforward and uses a calibrated power meter to measure the power of the pump beam before ($P_{in}$) and after the sample ($P_{out}$) outside the cryostat. To enlarge the transmission loss, we used for these measurements another crystal with a length of $L = 3.91$ mm. If $F_{rep}$ denotes the repetition rate and $T_p$ the pulse duration of the laser pulse (we assume a rectangular shape with $T_p = 3$ ps as a good approximation to the actual pulse) and a Gaussian spatial profile with beam waist $w_0$ the intensity inside at the entrance surface inside the sample is given by

$$I(r,z) = I_0 \left(\frac{w_0}{w(z)}\right)^2 \exp\left(\frac{-2r^2}{w(z)^2}\right) . \quad (1.34)$$

The relation between the average laser power and the peak intensity is

$$I_0 = \frac{2P_0}{\pi w_0^2} \frac{1}{T_P F_{rep}} \text{ and } w(z) = w_0 \sqrt{1 + \left(\frac{z}{z_R}\right)^2}. \tag{1.35}$$

With $P_0 = T_W (1 - R_{IR}) P_{in}$, whereby $T_W$ is the total transmission of the cryostat windows on one side and $R_{IR}$ denotes the reflectivity of the sample at the pump wavelength, the Rayleigh length is given by

$$z_R = \frac{\pi w_0^2 n}{\lambda}. \tag{1.36}$$

The beam waist $w_0$ is related to the beam diameter $D_{FWHM}$ at half intensity by

$$w_0 = \frac{D_{FWHM}}{\sqrt{2 \ln 2}}. \tag{1.37}$$

The beam diameter has been measured as $D = (125 \pm 5)$ μm giving a $w_0 = 106$ μm.

The intensity at the exit surface can be calculated from the two-photon absorption law (with two-photon absorption coefficient $\alpha_2$) by

$$\frac{dI_{IR}}{dz} = -\alpha_2 I_{IR}^2. \tag{38}$$

Neglecting any one or three photon absorption the integration (sample length $L$) gives for the intensity distribution at the crystal exit surface of the IR beam

$$I_{IR}(z=L,r) = \frac{I_0 \exp\left(\frac{-2r^2}{w_0^2}\right)}{1 + \alpha_2 L I_0 \exp\left(\frac{-2r^2}{w_0^2}\right)}. \tag{1.39}$$

The corresponding power $P_L$ we get by integrating over the surface as

$$P_L = P_0 \frac{\ln(1 + \alpha_2 L \frac{2P_0}{\pi w_0^2} \frac{1}{T_P F_{rep}})}{\alpha_2 L \frac{2P_0}{\pi w_0^2} \frac{1}{T_P F_{rep}}} \tag{1.40}$$

which is related to the measured output power by $P_{out} = T_W (1 - R_{IR}) P_L$.

As example we show the detailed evaluation in case of the 3S resonance. We get three sets of data by measuring the transmission, i.e., the ratio of $T = \frac{P_{out}}{P_{in}}$ for (i) without the crystal (calibration) at 1150.4 nm, (ii) with the crystal but at a wavelength well outside the 3S resonance (at 1151.2 nm) and (iii) exactly at the 3S resonance at 1149.2 nm. The results are shown in Fig. B1 for laser powers from 0.1 mW to 100 mW.

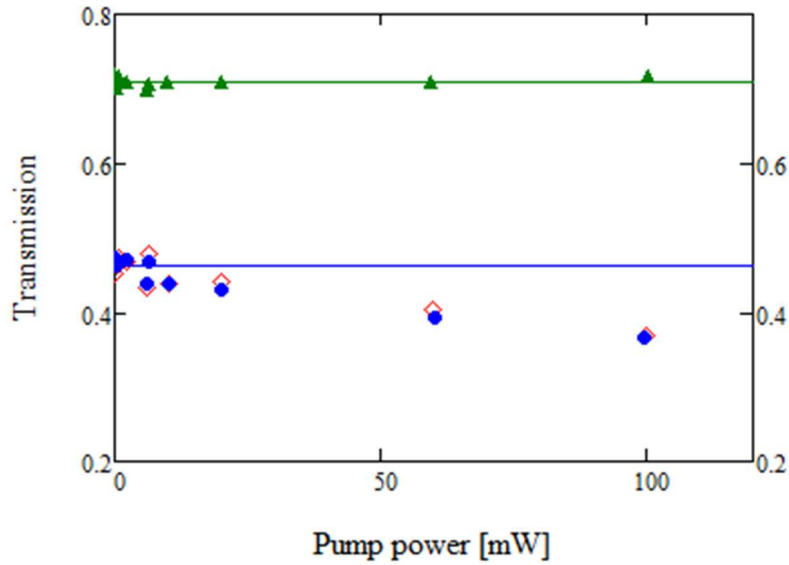

Fig. B1: Results of the transmission measurements of the 3S exciton. Green triangles: case (i), red open diamonds: case (ii) and blue dots: case (iii). The errors are about 0.4%. The full green and blue lines give the transmissions $T_0$ extrapolated to zero laser power. Note that in cases (ii) and (iii) we obtained identical values.

While the transmission in case (i) is power independent as expected for quartz windows, they show a pronounced non-linear behavior in the Cu$_2$O crystal, both in and out the resonance, whereby the intensity loss is slightly larger in resonance.

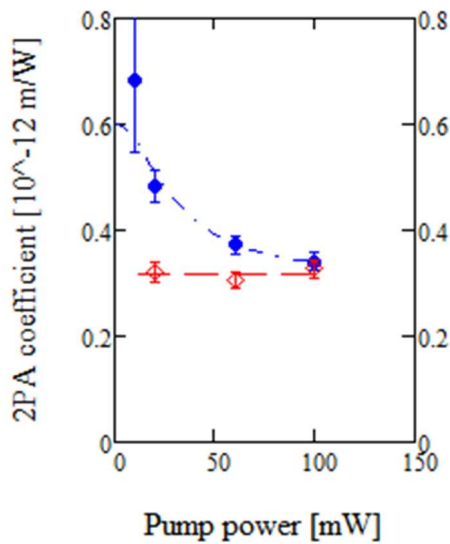

Fig. B2: 2PA coefficients for the cases (ii) (open red diamonds) and (iii) blue full diamonds as function of pump power. While out of resonance the coefficient is constant, it strongly depends on laser power in the resonant case (iii).

One can describe the situation by assuming an additional background two-photon absorption (2PA) given by a coefficient $\alpha_{2bg}$ .which can be induced by defects that lift the cubic symmetry of Cu2O, as was also observed in time-resolved 2PA [52]. It can be assumed to be wavelength independent, at least for the small difference in laser wavelength in and out resonance. Then the two-photon coefficient for resonance is just the sum of $\alpha_{2bg}$ and the resonant 2PA for the 3S exciton $\alpha_{23S}$.

By solving the non-linear Eq. (1.40) for each pump power, we obtain for the two cases (ii) and (iii) the 2PA coefficients as shown in Fig. B2.

While in the out of resonance case (ii) the coefficient does not depend on laser power, in the resonant case (iii) it is strongly power dependent and seems to saturate at high pump powers. As origin of this saturation one can straightforwardly identify the saturation of the exciton density by the Rydberg blockade effect which is given by an equation of the type

$$\alpha_{2nS}(P) = \frac{\alpha_{20nS}}{1 + \left(\frac{P}{P_{sat}}\right)^2}, \tag{1.41}$$

where $P_{sat}$ denotes the saturation power. As shown by the dash-dotted line in Fig. 2, the data can be quite well described by Eq. (1.41) with $\alpha_{203S} = (2.83 \pm 0.61) \cdot 10^{-12}$ m/W and $P_{sat} = (29.9 \pm 3.3)$ mW.

In our SHG experiments (thin sample) we need to know the density of the 3S excitons from the 2PA for very small distances from the entrance surface of the crystal. In this case, one can neglect the reduction of the IR laser power by 2PA and directly use Eq. (38) to calculate the number of absorbed photons by a single pulse giving as initial exciton density

$$\rho_{03S} = \frac{T_P}{2E_{IR}} \alpha_{23S}\left(\sqrt{T_0} P_{in}\right)\left(\frac{2}{\pi w_0^2}\frac{1}{T_P F_{rep}}\right)^2 T_0 P_{in}^2 \tag{42}$$

with the laser pulse width $T_P = 3$ ps. The resulting densities shown in Fig. B3. At the highest laser powers they are of the order of $5 \cdot 10^{15}$ cm$^{-3}$. At powers below 10 mW the density depends quadratically on the input laser power $P_{in}$ and can be described by the law $\rho_{3S} = C_{3S} P_{in}^2$ (with $C_{3S} = (4.36 \pm 0.79)$ mW$^{-2}$μm$^{-3}$ as shown by the dotted line in Fig.3.

Applying Eq. (7) we get for the constant $C_{PS} = (134.5 \pm 28.2)$ mW$^{-2}$μm$^{-3}$.

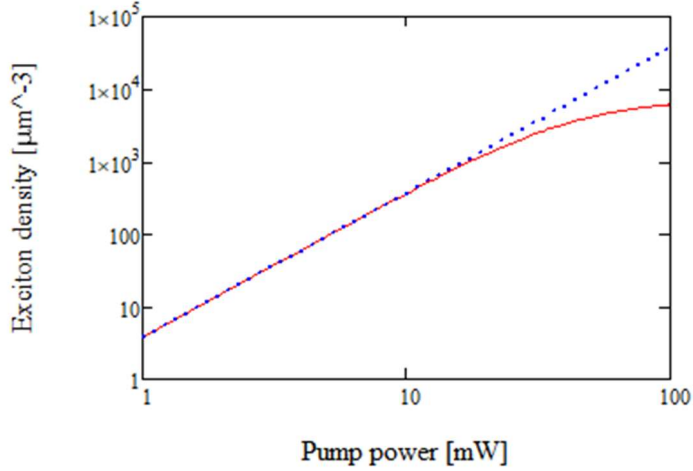

Fig.B3: power dependence of the 3S exciton density as obtained from the 2PA analysis (full red line). The dotted blue line gives the low power approximation Eq. (7).

**Appendix C: Analysis of the experiments at low powers**

In this Appendix we give the details of the evaluation of the data with the model of "absorption followed by emission" of the SHG process developed in Section IV. Here we used only experiments with so low laser power that the exciton density is below the Le Roy density (Fig.. The energy shifts are plotted in Fig. C1a. For $n = 3, 4, 5$ the fits show the expected square dependence on pump power, but for $n = 6, 7$ there already comes out a saturation behavior which is due to the blockade effect. To enhance the visibility of the blockade also for the small quantum numbers, we did plot the ratio of the peak areas to the square of the pump power in panel b, because any non-linearity shows up in such a plot. Indeed, the reduction of this ratio is clearly visible also in the low power range. By a least square minimalization the values for the parameters $a_1(n)$ from Eq. (11) and $V_{BL}(n)$ from Eq. (8) shown in Fig. 9 are obtained.

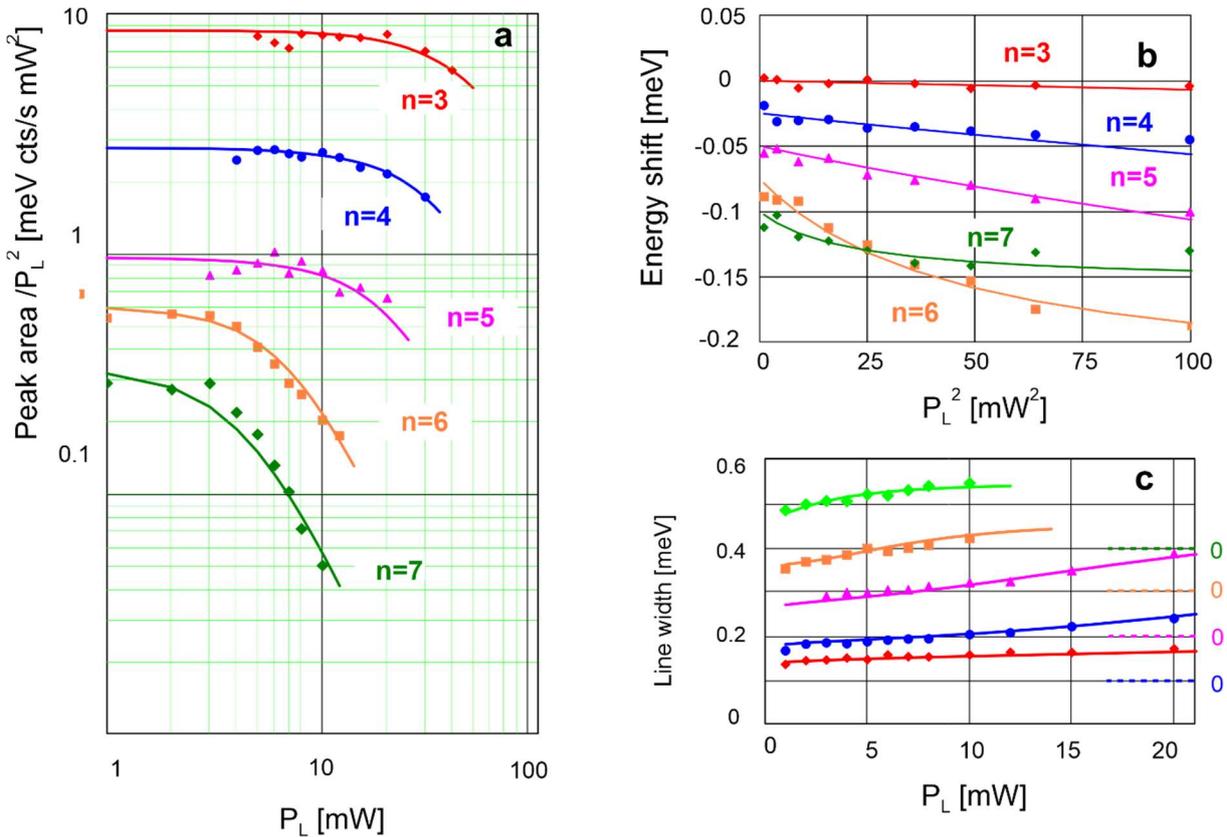

Fig. C1: Analysis of the peak intensity areas (panel a), energy shifts (panel b), linewidths (panel c) and of the experiments at the low laser powers with the absorption-followed-by-emission model from Section IVA and B. Points are the experimental data, the lines are the optimal fits with the theory. Note the different scales of abscissa and ordinates.